\documentclass[conference]{IEEEtran}
\IEEEoverridecommandlockouts

\usepackage{cite}
\usepackage{amsmath,amssymb,amsfonts}
\usepackage{graphicx}
\usepackage{textcomp}
\usepackage{xcolor}

\usepackage{tabularx}
\usepackage{adjustbox}
\usepackage[normalem]{ulem}
\usepackage{multirow}
\usepackage{booktabs}
\usepackage[framemethod=TikZ]{mdframed}
\usepackage{amsmath,amsfonts}
\usepackage{algorithm}
\usepackage{algpseudocode}
\usepackage{booktabs} 
\usepackage{listings}
\usepackage{verbatim}
\def\BibTeX{{\rm B\kern-.05em{\sc i\kern-.025em b}\kern-.08em
    T\kern-.1667em\lower.7ex\hbox{E}\kern-.125emX}}
\begin{document}

%\title{An XAI Service Framework for Analyzing the Quality Attributes of Open Vision Models Under Adversarial Attacks}
% \title{Cloud-Based XAI Framework for Model Quality Attributes Evaluation Under Adversarial Attacks}
\title{Cloud-based XAI Services for Assessing Open Repository Models Under Adversarial Attacks}

%Integrad cloud service for evaluating open repository models
%first class entities (data with augmentation, models, XAI library )
%introduction: motivating 
% how adopted with minimal efforts

% \author{\IEEEauthorblockN{Anonymous}
% \IEEEauthorblockA{\textit{Anonymous dept. name of organization} \\
% \textit{Anonymous name of organization}\\
% City, Country \\
% email address}}

\author{\IEEEauthorblockN{Zerui Wang}
\IEEEauthorblockA{\textit{Department of Electrical and Computer Engineering} \\
\textit{Concordia University}\\
Montréal, Québec, Canada \\
zerui.wang@concordia.ca}
\and
\IEEEauthorblockN{Yan Liu}
\IEEEauthorblockA{\textit{Department of Electrical and Computer Engineering} \\
\textit{Concordia University}\\
Montréal, Québec, Canada \\
yan.liu@concordia.ca}
}

\maketitle

\begin{abstract}
The opacity of AI models necessitates both validation and evaluation before their integration into services. To investigate these models, explainable AI (XAI) employs methods that elucidate the relationship between input features and output predictions. The operations of XAI extend beyond the execution of a single algorithm, involving a series of activities that include preprocessing data, adjusting XAI to align with model parameters, invoking the model to generate predictions, and summarizing the XAI results. Adversarial attacks are well-known threats that aim to mislead AI models. The assessment complexity, especially for XAI, increases when open-source AI models are subject to adversarial attacks due to various combinations. To automate the numerous entities and tasks involved in XAI-based assessments, we propose a cloud-based service framework that encapsulates computing components as microservices and organizes assessment tasks into pipelines. The current XAI tools are not inherently service-oriented. This framework also integrates open XAI tool libraries as part of the pipeline composition. We demonstrate the application of XAI services for assessing five quality attributes of AI models: (1) computational cost, (2) performance, (3) robustness, (4) explanation deviation, and (5) explanation resilience across computer vision and tabular cases. The service framework generates aggregated analysis that showcases the quality attributes for more than a hundred combination scenarios.
\end{abstract}


\section{Introduction}
\label{chap:introduction}
Artificial intelligence models are increasingly accessible through the open community \cite{huggingface} and facilitate advancements in software applications across various domains. 
These advancements, represented by innovations in deep-learning models such as ViT, ConvNeXt, CVT, Swin Transformers, SegFormer, and ResNet \cite{resnet,vit,swin,segformer,convnext,cvt}, underscore the rapid evolution of AI capabilities.
However, the integration of these models into software applications necessitates a strict evaluation \cite{runtime-based} of their quality attributes.
A noted gap in current practices is the absence of a comprehensive service framework that facilitates an understanding of model explainability \cite{doshi2017towards,lopes2022xai}. 

The importance of explainable AI (XAI) has been increasingly recognized \cite{gunning2019darpa}, driven by the need to build trust and ensure fairness within AI systems. XAI techniques, which aim to make the decision-making processes of AI models transparent \cite{arrieta2020explainable}, are essential in AI-enabled applications \cite{van2022explainable}.
Additionally, adversarial attacks target vulnerabilities of AI models.
They are modifications to input data that are less visible to humans but can make AI models give incorrect inferences or predictions \cite{moosavi2017universal,benchmarkingperturbations}. 
The rising threats \cite{zhang2019adversarial} targeting security-sensitive models introduce significant challenges to the deployment of AI in software services.
Therefore, effective software quality assurance requires a comparative analysis to guide the development and refinement of AI models, ensuring their robustness and explainability across diverse applications.

 The integration of XAI techniques into AI models introduces additional computational layers \cite{computationalcost}. The operational complexity of XAI evaluation \cite{xaiprocess} arises from multiple factors of managing diverse data types, integrating explanation methods with various AI models, evaluating explanations, and summarizing the results. These factors lead to evaluation scenarios that require dozens to hundreds of experiments. In addition, the complexity of assessment is further scaled by the product of multiple kinds of adversarial attacks. 
Comparative analysis between XAI and adversarial attacks increases the evaluation workload by necessitating the exploration of interactions between models and explanations under various adversarial conditions.
 
The goal is to address these complexities by automating the evaluation pipeline. 
We propose a cloud-based service framework that encapsulates computing components as microservices and organizes assessment tasks into pipelines. 
This framework also integrates open XAI tool libraries, which are originally not inherently service-oriented, as part of the pipeline composition. 

In scenario studies, we assess six vision models against three types
of adversarial attacks employing five XAI methods,
resulting in a total of ninety distinct combinations. 
We evaluate three transformer-based tabular models
using two XAI methods across three datasets, resulting in
eighteen combinations. We set pipelines to assess quality attributes for every combination scenario, including
(1) computational cost, (2) performance, (3) robustness, (4) explanation deviation, and (5) explanation resilience. 
We assessed these attributes across a range of model and XAI method combinations.
The evaluation results indicate that higher explanation deviation requires more computational costs.
We demonstrate the impact of adversarial attacks on model performance and their explanation.

\section{Background and Related Work}\label{chap:background}
In this section, we introduce the XAI methods. We review the current XAI tools and frameworks. Then, the taxonomy of adversarial attacks.

\subsection{Introduction of XAI Methods}

% XAI methods can be divided into interpretable model development and post-hoc explanation methods \cite{xaiprocess}. 
% The first option is to build models with human-understandable structures. 
% Decision tree algorithms \cite{charbuty2021classification} are capable of determining the conditions that cause a model to switch branches and make a decision. Conversely, post-hoc explanation methods \cite{xaiprocess} are designed to explain pre-trained models.

Mainstream post-hoc XAI methodologies can be categorized into two distinct types \cite{xaiprocess}: model-specific and model-agnostic methods. Model-specific methods \cite{Grad-cam,Grad-cam++,HiResCAM,xgradcam,layercam} derive feature importance values from the internal parameters of the model itself, offering insights directly linked to the model's internal mechanisms. In contrast, model-agnostic methods \cite{lundberg2017unified} establish feature importance by analyzing the relationship between the model's inputs and outputs without requiring the model's internal parameters and layers.

\begin{figure}[h]
\centering
\includegraphics[width=0.9\linewidth]{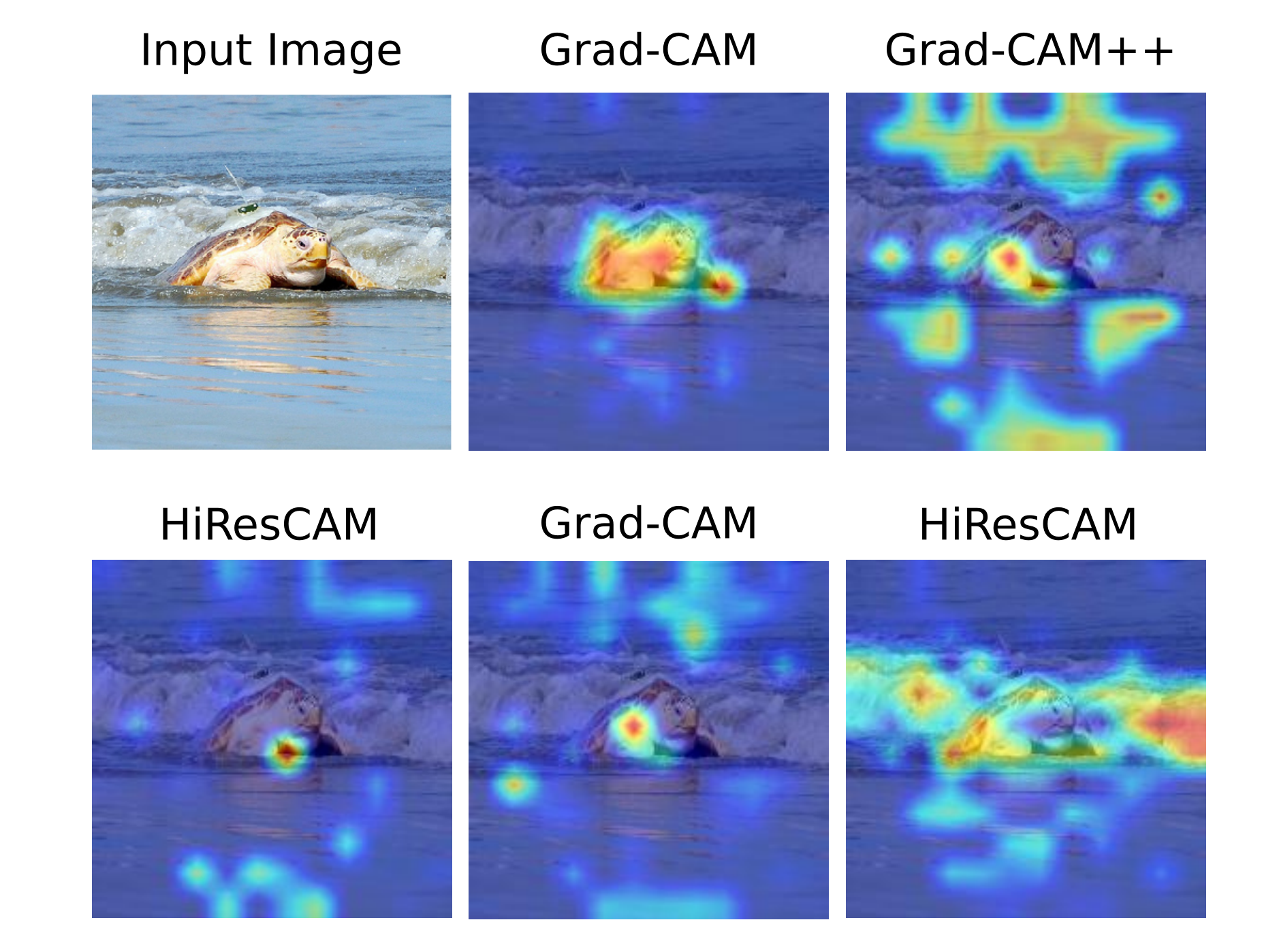}
\caption{Five CAM-based Visual Explanations from Vision Transformer Model with One Image Example.}
\label{fig:example_cam}
\end{figure}

XAI in computer vision often involves generating visual explanations for the decisions made by AI models. The model-specific methods are computationally efficient, rather than the model-agnostic methods \cite{abhishek2022attribution}. 
For demonstration, Figure \ref{fig:example_cam} presents selected examples derived from the Vision Transformer model \cite{vit} through our XAI service framework.
Significant variations in results are observed from the same input data. These results indicate the need for establishing systematic XAI evaluations within the XAI service framework, which is notably absent in previous tools and frameworks.

Grad-CAM \cite{Grad-cam} utilizes the gradients of the target label flowing into the final convolutional layer to produce a coarse localization map highlighting important regions for prediction. According to an optimization work, Grad-CAM++ \cite{Grad-cam++} extends Grad-CAM by considering the weight of each pixel in the feature maps, allowing better handling of images with multiple occurrences of the same object. HiResCAM \cite{HiResCAM} generates high-resolution class activation maps, allowing for finer detailed visual explanations with higher computational cost. XGrad-CAM \cite{xgradcam} focuses on increasing the linearity of the saliency maps, providing improvements. LayerCAM \cite{layercam} introduces a layer-wise relevance propagation mechanism, which enables the feature importance across different network layers for more details.

\begin{figure}[h]
\centering
\includegraphics[width=0.9\linewidth]{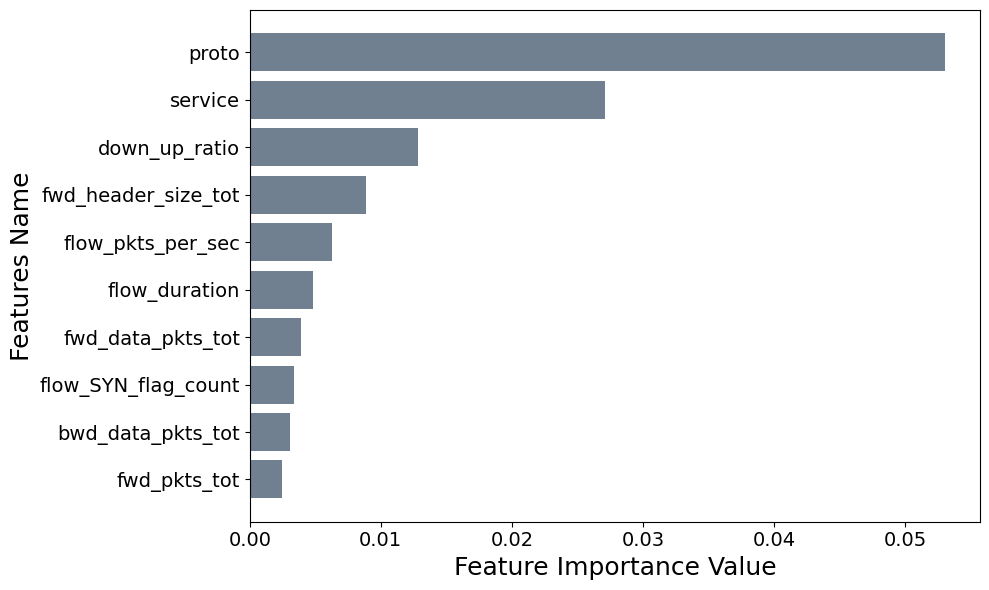}
\caption{Top 10 out of 83 SHAP Feature Importance Explanations from FT Transformer on RT-IoT Cybersecurity Threats Dataset.}
\label{fig:example_shap}
\end{figure}

We also apply XAI methods involving structured data within the tabular models. Mean Centroid Prediff \cite{Trustworthy} and SHAP (SHapley Additive exPlanations) \cite{lundberg2017unified} stand out for widespread acceptance as the baseline. These methods quantify feature importance by assessing the impact of masking each feature on model output. The methods provide a score for each feature, indicating its contribution to model predictions.
Figure \ref{fig:example_shap} illustrates an example of the top ten out of eighty-three global feature importance explanation from the RT-IoT2022 cybersecurity classification dataset \cite{rt-iot}.
We carry out XAI to determine the impact of the network logs features on the classification of threats.
We provide data-driven explanation deviation metrics to evaluate XAI methods in the section \ref{chap:results_discussion} scenario studies.

\subsection{XAI Tools Libraries and Frameworks}
\label{sec:reviewframework}

The role of XAI in providing feature contribution explanations establishes trust in AI models \cite{Trustworthy}. 
Before introducing XAI as a service, we review the published tools and libraries. The Explainability 360 toolkit by IBM~\cite{aix360} integrates their explanation techniques within the toolkit package. Microsoft's InterpretML~\cite{interpretml} offers support for eight tabular models' XAI approaches. The recent framework OmniXAI~\cite{yang2022omnixai} offers a broad range of techniques for XAI. 
However, based on our tests and usages, the existing XAI frameworks have the following limitations:

\textbf{Expertise-Dependent Usage:} The use of these tools often demands specific XAI knowledge expertise, which limits accessibility \cite{palacio2021xai} to software engineers.

\textbf{Restricted Methods Support:} Each library support limited numbers of different XAI methods \cite{aix360,interpretml,yang2022omnixai}.The disparity in the number of methodologies supported by these tools complicates the selection process and necessitates additional preprocessing.

\textbf{Evaluation Procedures Deficiency:} The comprehensive study \cite{saliency} that performed quantitative evaluations on saliency methods is relevant. However, the current XAI tools lack standardized procedures for evaluating explanation results, which limits the selection and enhancement of these XAI tools~\cite{Zhang2021eva,lopes2022xai}. 

\textbf{Cloud Service Support Limitations:} Explicit support for cloud AI services is rare among these tools, affecting their applicability across cloud service environments.

These limitations underline the need for a service architecture that addresses these gaps, making XAI accessible and effective for diverse AI-enabled applications.

\subsection{Adversarial Attacks Types}

\begin{figure}[h]
\centering
\includegraphics[width=\linewidth]{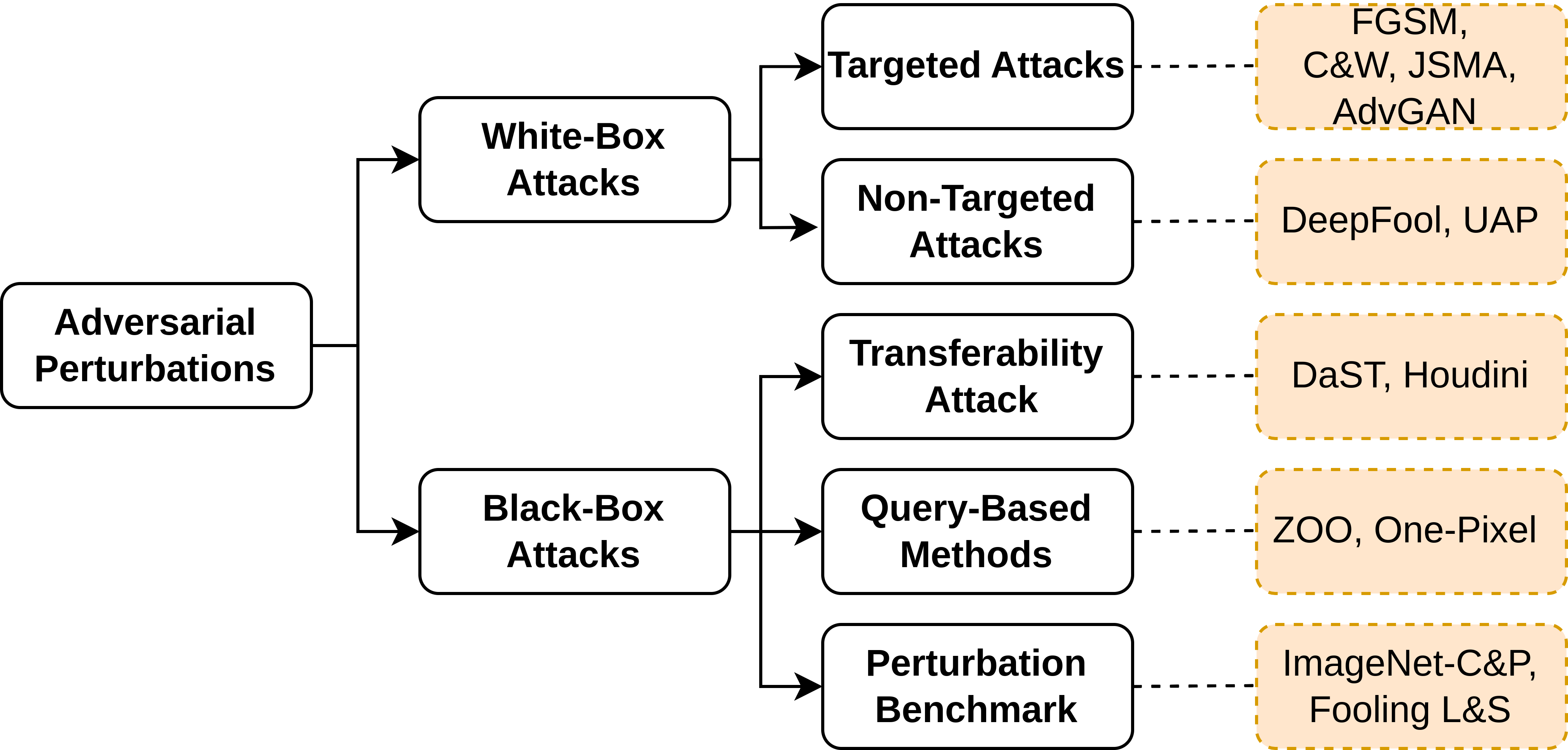}
\caption[Taxonomy of Adversarial Attacks]{Taxonomy of Adversarial Attacks. {\normalfont\footnotesize References: FGSM \cite{FGSM}, C\&W \cite{C&W}, JSMA \cite{JSMA}, AdvGAN \cite{AdvGAN}, DeepFool \cite{DeepFool}, UAP \cite{UAP}, DaST \cite{DaST}, Houdini \cite{Houdini}, ZOO \cite{ZOO}, One-Pixel \cite{One-Pixel}, ImageNet-C \cite{imagenetC}, ImageNet-P \cite{imagenetC}, Fooling LIME and SHAP \cite{fooling}.}}
\label{fig:taxonomy}
\end{figure}

Adversarial attacks are designed to mislead models, and pose challenges to AI system security \cite{AdvGAN}.
The adversarial attacks are briefly categorized into white-box and black-box attacks, each with distinct methodologies and implications. The taxonomy of adversarial attacks with example works are listed in Figure \ref{fig:taxonomy}.

%In white-box attacks, the attacker needs certain knowledge of the model.
%They aim to manipulate the model to produce incorrect output.
%In black-box attacks, the focus shifts to developing attacks based on transferability, query, and perturbations. 

In white-box attacks, the attacker needs certain knowledge of the model, including its architecture and parameters. Targeted attacks \cite{FGSM,C&W,JSMA,AdvGAN} aim to manipulate the model to produce a specific, incorrect output. Non-targeted attacks \cite{DeepFool,UAP} aim to generate any incorrect response from the model. 
% DeepFool \cite{DeepFool} is an adversarial attack method that computes minimal perturbations to shift individual inputs across the decision boundary of a model, assuming a locally linear boundary. Universal Adversarial Perturbations (UAP) \cite{UAP} aims to create a single, broadly effective perturbation that can mislead a model on a wide range of inputs, highlighting the general susceptibility of models to such attacks.

In black-box attacks, where the attacker has no information about the model's internals, the focus shifts to transferability attacks, query-based methods, and perturbation benchmarks. The perturbation benchmarks refer to algorithmically induced modifications applied to the data with the intent to mislead the model. The benchmark modifications can affect models and further be employed to measure the model's robustness \cite{imagenetC}.
The study \cite{imagenetC} presents corruption and perturbation as two methodologies for modifying images. Corruption typically refers to modifications that simulate natural or environmental degradation. Perturbation denotes generated changes in sequence.
We merge these two groups algorithmically modify methods in the study \cite{imagenetC}, under the broader category of adversarial perturbations.
In the assessing scenarios, we adopt the ImageNet-C benchmark \cite{imagenetC}.
The benchmark \cite{imagenetC} contains fifteen types of algorithmically generated corruptions from categories such as noise, blur, weather, and digital. The weather category contains corruptions that demonstrate an excessive diversity. In the assessment scenarios, we select a representative algorithm from each of the three main categories. We apply nine distinct corruption perturbations across three levels of severity.
The Fooling LIME and SHAP \cite{fooling} method is a recent study using data perturbation to attack LIME \cite{LIME} and SHAP \cite{lundberg2017unified}. 
We also develop and launch multiple levels of perturbation attacks to the numerical features in selected tabular datasets for scenarios.

\section{Overview of AI System's Quality Attributes }
\label{sec:Recapping}

AI-based system refers to a software system that adopts the AI models as components \cite{felderer2021quality}. Software quality is the ability of a software product to meet requirements during its operation in designated conditions \cite{kurtanovic2017automatically}. Software quality assurance involves an evaluation process of how well a software product satisfies its stated needs \cite{naik2011software}. 
However, the AI models are distinct from conventional software components. Their core characteristic is being data-centric \cite{8804457}. Additionally, these components are dynamic and continually evolving as they are exposed to new data over time \cite{8804457}. Inadequate, inaccurate, or unsuitable training data can result in unreliable models and biased decisions \cite{budach2022effects}. To address quality assurance, the study \cite{icsexaiport} presents the concept of early adoption of XAI into model development.

%We identify and summarize the key quality attributes that are critical for model evaluation. 
We define the quality attributes of AI models and the explanations produced by XAI analysis in terms of \texttt{model performance} and \texttt{explanation deviation}.
Under adversarial attacks, these quality attributes are further specified as \texttt{model robustness} and \texttt{explanation resilience}. 
Additionally, \texttt{computation cost} is considered in relation to deployment. 
The metrics established for these quality attributes encapsulate the combined states of AI models, XAI methods, adversarial attacks, and datasets. 
The results can be visualized using a radar chart, where each quality attribute is transformed into a normalized value.

\textbf{Computational Cost.}
Computational cost measures the resources required to execute an algorithm. The computational cost includes measuring the utilization of CPU, GPU, and Memory. 
We record the computational cost attribute that encompasses the runtime and energy consumption of AI models and XAI techniques. 
These metrics are measured in seconds (s) for runtime and watt-hours (Wh) for energy consumption.
The CodeCarbon library \cite{codecarbon} supports the program to track resource utilization by monitoring hardware-specific parameters. 
In addition, the CodeCarbon estimates the environmental carbon footprints based on energy consumption.

%The complexity of machine learning has experienced significant growth \cite{computationalcost}. 
%The CUDA-enabled GPUs accelerate AI-enabled applications \cite{runtime-based} is the main fraction in the consumption. 

\textbf{Model Performance. }
The evaluation of model performance contains a range of metrics. The typical metrics are Top-N accuracy metrics\cite{cremonesi2010performance}, precision, recall, and F1 score \cite{presicionrecall}. In the multi-class context, we aggregate True Positives (TP), False Positives (FP), True Negatives (TN), and False Negatives (FN) across all classes. In addition, the Area Under the Receiver Operating Characteristics Curve (AUC-ROC) is also a calibrating metric to assess a model's ability to distinguish between classes \cite{ROC}. 
%This service framework provides a multifaceted approach to performance evaluations that support researchers and practitioners in the development and refinement of AI system performance.

\textbf{Model Robustness.}
Robustness evaluation measures a model’s performance degradation under specific adversarial conditions.
For computer vision models, the use of an image adversarial perturbation benchmark \cite{imagenetC} assesses the model under adversarial conditions. Mean Corruption Error (mCE) represents the model robustness in a previous study \cite{imagenetC}.
This metric aggregates the normalized error rates for a model across corruption types and their respective severity levels. The formulation for mCE \cite{imagenetC} is detailed below:

\begin{equation}
mCE = \frac{1}{S_c} \sum_{s=1}^{S_c} \left( \frac{E_s^c(f)}{E_s^c(f_{\text{ref}})} \right)
\end{equation}

Here, \(f\) denotes the model under evaluation. For corruption type \(c\), \(S_c\) indicates the severity levels. The variable \(E_s^c(f)\) is the error rate of the model \(f\), while \(E_s^c(f_{\text{ref}})\) specifies the error rate of a reference model, such as Alexnet according to the study \cite{imagenetC}. 

Instead of comparing with a reference model, we set Kolmogorov-Smirnov (K-S) statistic \cite{KolmogorovSmirnov} \(D_{ks}\) as a quantifiable metric. It represents the comparison of the distribution of the model outputs between two datasets:

\begin{equation}
D_{ks} = \sup_{X} \left| F(X_{\text{orig}}) - F(X_{\text{adv}}) \right|
\end{equation}

Here, \(F(X_{\text{orig}})\) and \(F(X_{\text{adv}})\) symbolize the model probabilities distribution of the original and adversarial datasets, respectively. The usage of \(\sup_{x}\) targets the maximum divergence between these two model inference distributions.
By incorporating the K-S statistic, we redefine the robustness against adversarial attacks. Equation \ref{eq:robustness} represents the assessment. A smaller value indicates better model robustness:

\begin{equation}
Robustness =  \frac{1}{S_c} \sum_{s=1}^{S_c} D_{ks}
\label{eq:robustness}
\end{equation}

For the tabular model, we generate feature perturbations to datasets. Specifically, in the scenario, we make a random perturbation to the feature with the numerical severity factor. Similar to our approach, the adversarial attack \cite{fooling} introduces a designed biased perturbation, instead of our random perturbation, to attack the SHAP \cite{lundberg2017unified} and the LIME \cite{LIME} methods. The robustness Equation \ref{eq:robustness} can be used in tabular cases.

\begin{figure*}[ht]
\centering
\includegraphics[width=\linewidth]{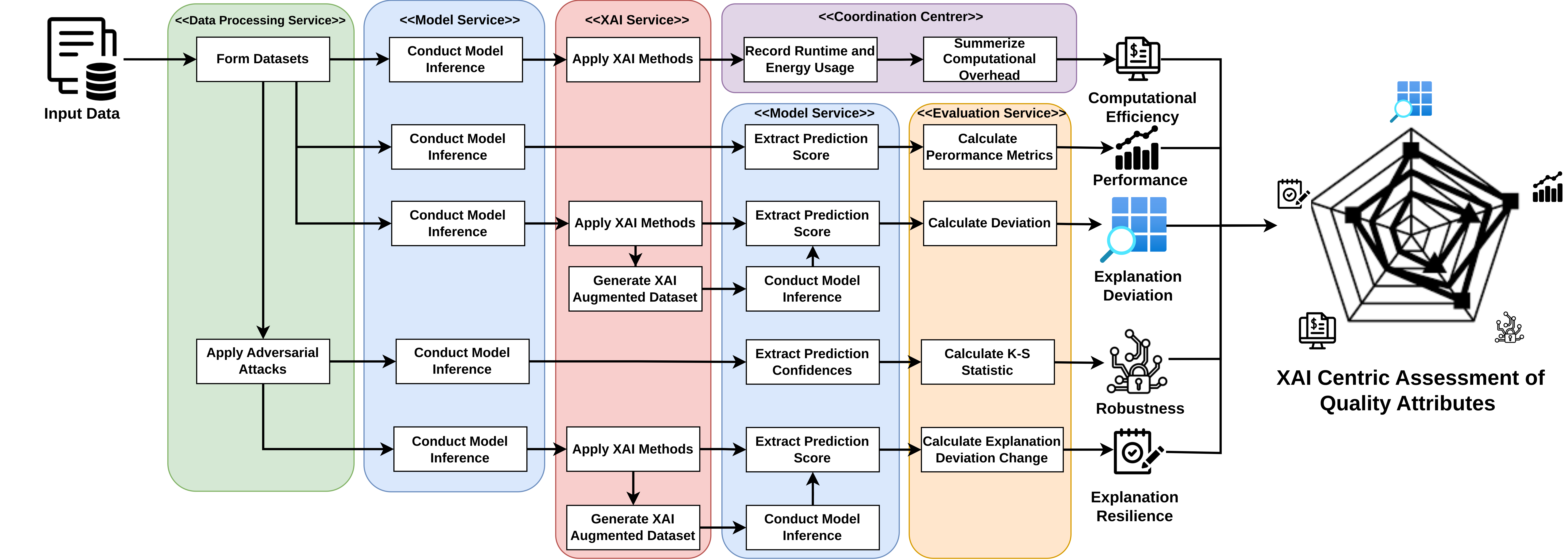}
\caption{Assessment Pipelines for Open-source AI Model Quality Attributes.}
\label{fig:evaluation_process}
\end{figure*}

\textbf{Explanation Deviation.}
Explanation deviation is an assessment of the impact of the feature importance explanation on the model's outputs.
At its core, explanation deviation measures the discrepancy between a model's predictions when all features are considered versus when only those important features are emphasized.
Therefore, measuring explanation deviation is a means to validate the actual influence of presumed features on the model's outputs. 

In vision scenarios, the saliency map is applied to assess how important each pixel is to a model's output, creating visual importance maps. 
Grad-CAM \cite{Grad-cam} is commonly accepted in CNN-based vision model \cite{resnet} explanation. However, whether these XAI perform as effectively for more transformer-based models remains under evaluation \cite{sobahi2022explainable}. 
Explanation deviation is determined by the prediction change score, which indicates the impact of the saliency map area on the model's predictions.
% For the detailed method implementation, we assess whether these saliency-highlighted areas dictate the model's classification predictions. 
% The function \(f\) represents the model being assessed.
The masked image is calculated as the element-wise multiplication of the original image \(X\) and the normalized, three-dimensional saliency mask \(M_{\text{norm 3D}}\).
Here, \(P\) is defined as the model's prediction class probability values:
% \begin{equation}
% X_{\text{masked}} = X \cdot M_{\text{norm 3D}}
% \end{equation}
\begin{equation}
Deviation =  1 - (P(X_{\text{orig}}) - P(X_{\text{orig}} \cdot M_{\text{norm 3D}}))
\end{equation}
A smaller change in probability value, then the deviation close to one.
Extending to the whole dataset, the overall explanation deviation is the statistical analysis of their median value. 
In tabular scenarios, the XAI method computes feature importance values for each data sample. 
% Feature importance values from different XAI methods are not numerically comparable \cite{xaiprocess} due to their varying XAI algorithms.
% However, feature importance values can be ranked in order, showing the sequence of features influencing model outcomes.
The explanation deviation is represented by evaluating the consistency in the feature importance order, according to the study \cite{xaiprocess}.
% The average Kendall tau distance \cite{xaiprocess} across feature importance orders can be computed. 

\textbf{Explanation Resilience.}
Resilience to adversarial attacks is measured by the difference in explanation deviation between non-adversarial and adversarial conditions.
Therefore, the explanation resilience is quantified by the Equation \ref{eq:explanation_resilience}:

\begin{equation}
\begin{split}
Resilience = & (P(X_{\text{orig}}) - P(X_{\text{orig}} \cdot M_{\text{norm 3D}})) \\
& - (P(X_{\text{adv}}) - P(X_{\text{adv}} \cdot M_{\text{norm 3D}}))
\end{split}
\label{eq:explanation_resilience}
\end{equation}

The resilience attribute indicates that the explanation deviation metric decreases for adversarial reasons. 
For tabular data, resilience can be calculated by subtracting the explanation deviation of adversarial data from the original data.
Comparing the original and perturbed situation allows for resilience assessment: The smaller the resilience, the better the explanation can resist the impact of adversarial attacks.

\section{Pipelines of XAI Centric Assessment of Quality Attributes}
\label{chap:workflow}

In this section, we define the pipeline for obtaining quality attributes. Then, we illustrate the cloud-based service architecture. The operational overhead for XAI service is compared with using the existing framework.

\subsection{Define Pipelines for Each Quality Attribute}
This study integrates five quality attribute evaluations. 
Figure \ref{fig:evaluation_process} shows the multiple parallel processes for accessing quality attributes.
This multifaceted assessment evaluates quality attributions under both original and adversarial conditions.

\textbf{Computational Cost Pipeline:} This pipeline records the computational resources used during model inferences and XAI method executions. The average resource consumption is logged by the \texttt{Coordination Center}, when processing a significant volume of data inputs. The record covers AI models, XAI methods and evaluations.

\textbf{Model Performance Pipeline:} To assess model performance, the pipeline begins with the datasets provided by the \texttt{Data Processing Microservice}. 
These datasets are sequentially fed into the \texttt{Models Microservice}, which encapsulates the pre-trained model.
The outcomes are extracted and persisted. The \texttt{Evaluation Microservice} derives performance metrics. 

\textbf{Explanation Deviation Pipeline:} This pipeline calculates explanation deviation. \texttt{XAI Methods Microservice} are applied to generate explanations. For computer vision tasks, the explanation results are fed back into the model to calculate the prediction confidence drop, thereby calculating explanation deviation. The prediction changes value within consistency metric \cite{xaiprocess} can also be directly derived from tabular models. 

\textbf{Robustness Pipeline:} To assess model performance decline under adversarial attacks, the pipeline begins with the \texttt{Data Processing Microservice} preparing perturbed data. 
The \texttt{Models Microservice} then processes both the original and perturbed datasets, yielding two sets of results.
The \texttt{Evaluation Microservice} measures shifts in the model's performance between the original and perturbed datasets.

\textbf{Explanation Resilience Pipeline:} Similar to the robustness pipeline but with a focus on XAI methods, this pipeline starts with the \texttt{Data Processing Microservice} preparing perturbed datasets.
These datasets are then subjected to selected models and XAI methods.
The \texttt{Evaluation Microservice} calculates the changes in explanation deviation.

The individual pipeline is configurable for AI models, XAI methods, and adversarial datasets. They can be composed into singular or combined quality attributes assessment scenarios. Their deployment and execution need a runtime service-oriented architecture, which is defined in the next subsection.

\subsection{Define Services Architecture for the Pipeline Configuration}

\begin{figure}[ht]
\centering
\includegraphics[width=\linewidth]{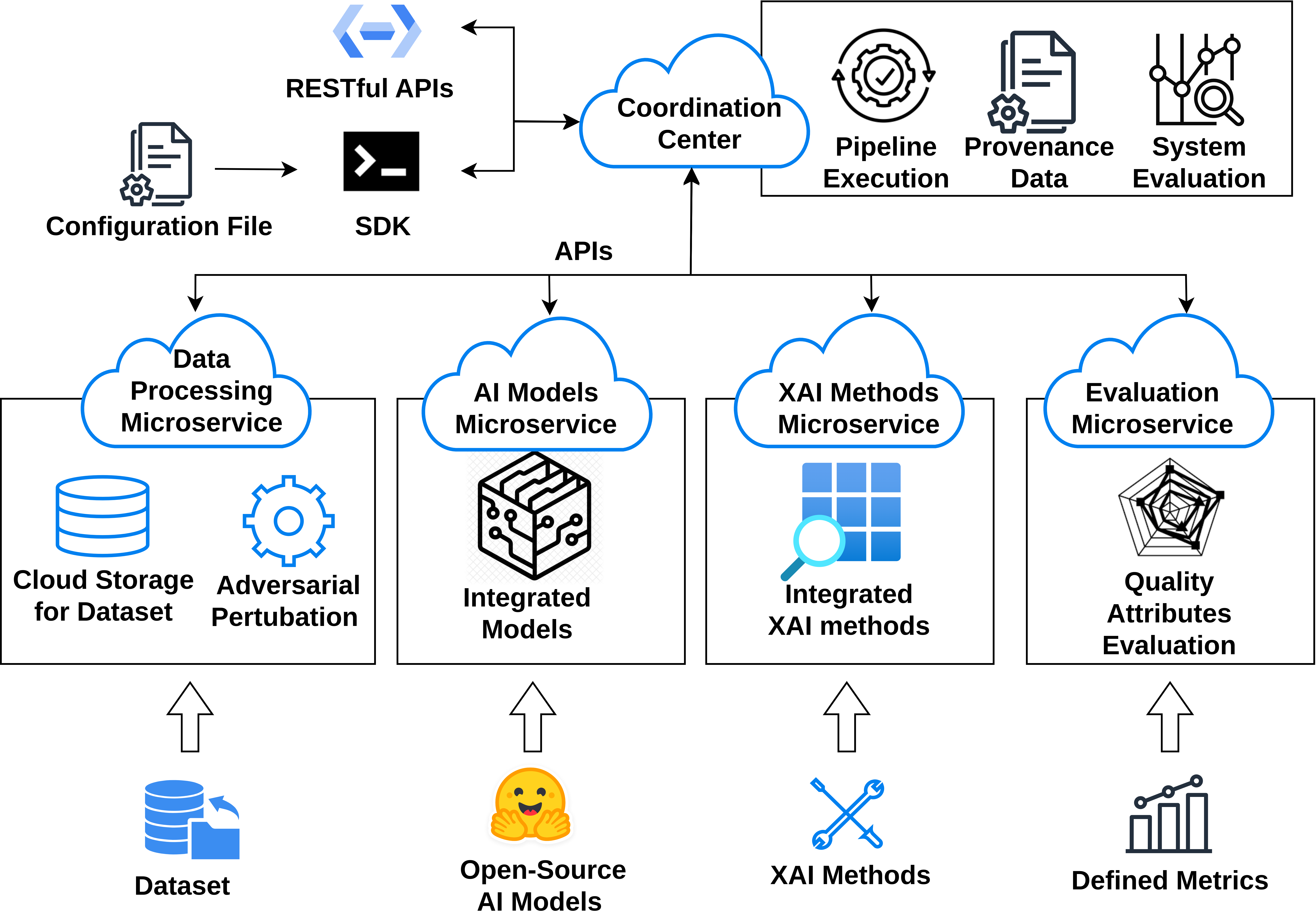}
\caption{Cloud-based XAI Service Architecture.}
\label{fig:servicearchitecture}
\end{figure}

We introduce a cloud service architecture designed to develop XAI assessment pipelines. This architecture enables analysis and comparison of various combinations of AI models, XAI methods, datasets, and adversarial attack approaches, as shown in Figure \ref{fig:servicearchitecture}.

\textbf{Coordination Center:} The microservice executes unit operations based on configuration, communicates with other microservices as per task setup, and records provenance data for transparency and reproducibility.

\textbf{Data Processing Microservice:} This microservice ensures data is correctly formatted and meets XAI algorithm requirements. It also applies adversarial attack conditions.
% to the original datasets.

\textbf{AI Model Microservice:} This service encapsulates and deploys pre-trained AI models, including open-source models from the HuggingFace Community, within its framework.

\textbf{XAI Method Microservice:} This service offers XAI that can generate explanations. XAI algorithms \cite{Trustworthy} and tool libraries  \cite{yang2022omnixai} can be encapsulated into the service. 

\textbf{Evaluation Microservice:} This service aggregates results and systematically evaluates defined quality attributes.

Collectively, these components create a cloud-based architecture supporting the complete evaluation process.
The architecture's flexibility allows for switching services to test various AI models or XAI methods, facilitating extensive investigation into numerous combinations.
Additionally, the deployed services are reusable across multiple pipelines.

\begin{figure}[h]
\centering
\begingroup
\footnotesize
\begin{verbatim}
"xai_config": {
  "base_url": "xaiport.ddns.net:8003", //"address"
  "datasets": {
    "t1024_gaussian_2": {     //"datasets id"
      "model_name": "resnet", //"model name"
      "algorithms": [
        "GradCAM", //"XAI methods name"
        "HiResCAM", 
        "GradCAMPlusPlus", 
        "XgradCAM", 
        "LayerCAM" 
      ]
    }
  }
}
\end{verbatim}
\endgroup
\caption{Sample JSON Template for XAI Methods Configuration}
\label{fig:xai_config}
\end{figure}

In addition, the JSON-based configuration template, as present in Figure \ref{fig:xai_config}, specifies how the service executes the pipeline according to the user's inputs. The template allows users to define the interaction between different services and customize the evaluation process to suit specific requirements. 
The execution of pipelines is through the use of coordination centers, each configured with its JSON-based configuration file. Upon receiving a configuration file, a coordination center systematically accesses the specified microservices to execute each pipeline step.

\subsection{Comparative Evaluation and Service Integration of Existing XAI Frameworks}
The architecture enables the encapsulation and customisation of the external XAI algorithms and frameworks. 
XAI microservices can encapsulate not only XAI methods but also published libraries and frameworks. For instance, OmniXAI \cite{yang2022omnixai} aggregates enriched XAI methods for various data types. 
We import the OmniXAI package and employ the related functionalities to compute explanations in the XAI methods microservices. 
The rest of the required units can continue to adopt our service architecture.
To verify the integration of the external tool framework, We compare our XAI service and the recently published OmniXAI framework \cite{yang2022omnixai}. We focus on differences in task reproducibility, metrics, and ease of use.

\textbf{Reproducible:} Reproducible ensures that results can be reliably validated.
Other frameworks offer explanation methods but lack detailed provenance data recording.
Validating and reproducing an XAI process requires obtaining the original dataset and editing the source code.
This demands effort to reproduce the complex XAI process manually.
Our service framework makes the evaluation pipeline reproducible using provenance data.
Users can execute the pipeline with a single command using provenance data as the configuration file.

\textbf{Evaluation Metrics:} Other XAI frameworks do not provide evaluations for their methods. 
Without rigorous evaluation, the effectiveness of the explanations provided by different XAI remains undetermined. 
Our framework introduces the calculation of XAI consistency metrics \cite{xaiprocess}. 
We also define the XAI-centric assessment quality attributes.

\begin{table}[ht]
\centering
\caption{Operational Overhead Comparison Between XAI Service and Tool Framework}
\label{tab:operational_comparison}
\footnotesize % Sets text to footnote size
\begin{tabularx}{\columnwidth}{|p{0.15\columnwidth}|p{0.33\columnwidth}|p{0.37\columnwidth}|}
\hline
\textbf{Steps} & \textbf{Our Service} & \textbf{OmniXAI\cite{yang2022omnixai} Framework} \\
\hline
Data Preparation & Automate data upload with formatted templates. & Script data transformation and insertion. \\
\hline
Environment Setup & Automate environment setup through Docker container. & Manage dependency installation, library configuration, and compatibility. \\
\hline
Configuration & Utilize JSON for configuration with defined rules. & Configure interactions and dataset linkages manually.\\
\hline
Pipeline Execution & Execute pipeline with a single SDK command. & Load datasets, inferences, and compute XAI manually. \\
\hline
Results Analysis & Provide built-in metrics and visualization. & Summarize results with custom coding. \\
\hline
Adjustments & Support adjustments via JSON editing. & Modify code and restart processes manually. \\
\hline
\end{tabularx}
\end{table}

\textbf{Ease of Use:} Our framework provides a streamlined SDK interface that simplifies pipeline execution.
Table \ref{tab:operational_comparison} provides a described comparison of the operational steps.
The JSON-based configuration template allows users to select datasets, models, and XAI methods in a structured manner.
By automating several steps such as data preparation and environment setup, our service minimizes the operational overhead for systematic XAI processes.

\textbf{Result Values Comparison:} We compared the explanation deviation results of our service framework with those of existing tools using the ImageNet dataset \cite{ImageNet,imagenetC} test set, containing ten thousand images.
\begin{figure}[htbp] 
\centering 
\includegraphics[width=1\linewidth]{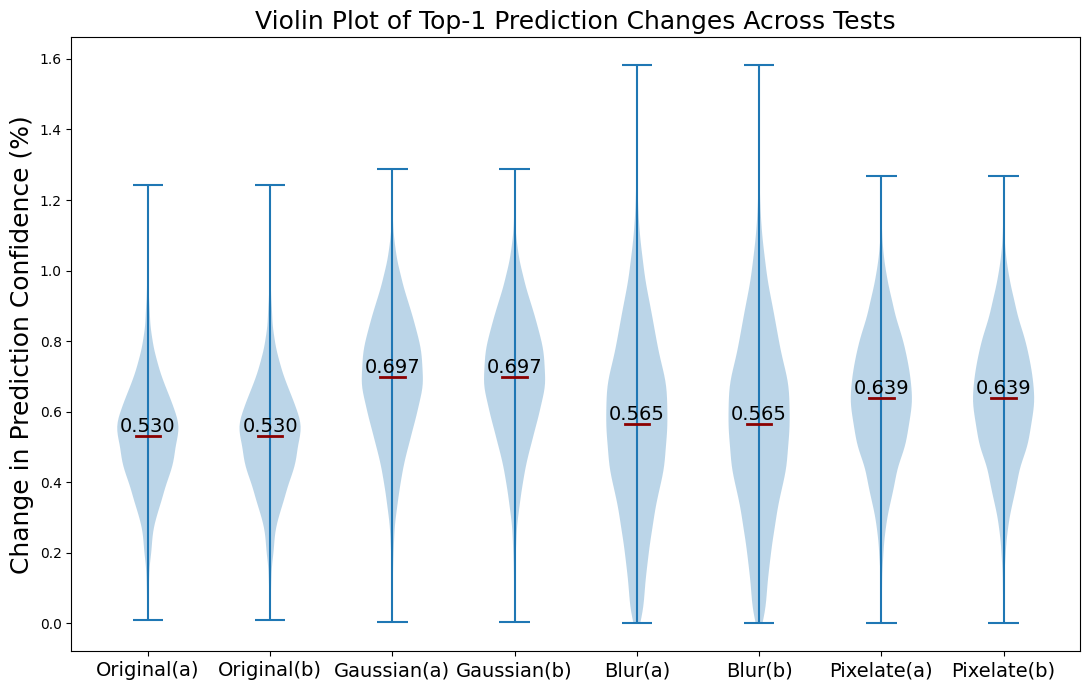} 
\caption[Explanation Deviation Analysis for XAI Service (a) Versus OmniXAI (b) Using GradCAM on ResNet]{Explanation Deviation Analysis for XAI Service (a) Versus OmniXAI (b) Using GradCAM on ResNet} 
\label{fig:omni_comparison} 
\end{figure}

Our XAI service includes multiple CAM-based methods \cite{Grad-cam,Grad-cam++,HiResCAM,xgradcam,layercam}. The recent tool framework OmniXAI \cite{yang2022omnixai} employs only Grad-CAM \cite{Grad-cam} as the XAI method for vision models. 
Figure \ref{fig:omni_comparison} presents a comparative analysis of the results from implementing Grad-CAM algorithms in our service framework versus OmniXAI.
The results show the same prediction confidence distribution. However, our XAI service framework significantly streamlines operations.

Additionally, our framework is containerized and suitable for cloud platform environments. The RESTful API design enables easy integration with external cloud model services.

% \textbf{Computational Cost:} The computational cost is determined by the algorithms employed. Different from other frameworks, our system tracks computing resources, time, and energy consumption using the CarbonCode library \cite{carboncode}. This feature assists software engineers in estimating the costs associated with deploying and evaluating AI systems.
\section{Assessment Scenarios and Quality Attribute Analysis}
\label{chap:results_discussion}

We introduce experimental scenarios to demonstrate the service's functionality in scenarios.
By running the pipelines defined by the service framework, we aim to investigate the research questions as follows: 
\begin{itemize}
    \item \textit{\textbf{RQ1}: Are the explanation deviation generated by XAI methods variable across models with different structures?} 
    \item \textit{\textbf{RQ2}: What is the relationship  between computational cost and explanation deviation in model-XAI combinations?} 
    \item \textit{\textbf{RQ3}: Considering the known impacts of adversarial perturbations on model performance metrics, how do these perturbations influence the explanation deviation?} 
\end{itemize}

The related source code and experimental results can be found in the GitHub repository \footnote{https://github.com/ZeruiW/XAIport}.
Experiments are conducted in a controlled environment to ensure consistency.
The experimental evaluation used a local setup with an Nvidia RTX 4090 GPU based on the AD102 graphics processor.
This GPU features a core clock speed of 2.52 GHz and 24 GB GDDR6X VRAM. All experiments were conducted on a Linux-6.2.0 system, Ubuntu 22.04 LTS, Python 3.8.18 with CUDA 12.1, and PyTorch 2.1.0 to leverage the GPU's capabilities.

\subsection{The Assessment Scenario of Vision Models}
\textbf{Process Configuration:}
Our dataset comprised 10,000 images from ImageNet \cite{ImageNet}, covering one thousand classes. 
We introduced three types of adversarial perturbations in the ImageNet-C \cite{imagenetC}: Gaussian Noise, Defocus Blur, and Pixelate, with each type occurring at three severity levels. 

\begin{table}[ht]
\centering
\caption{Selected State-of-the-Art Vision Models from the Huggingface Repository}
\label{tab:models_overview}
\begin{tabular}{lll}
\toprule
\textbf{Vision Model} & \textbf{Publisher} & \textbf{Huggingface Repository} \\
\midrule
ViT & Google & google/vit-large-patch32 \\
ConvNeXt & Meta & facebook/convnext-tiny \\
CVT & Microsoft & microsoft/cvt-13 \\
Swin & Microsoft & microsoft/swin-large-patch4 \\
SegFormer & Nvidia & nvidia/mit-b0 \\
ResNet & Microsoft & microsoft/resnet50 \\
\bottomrule
\end{tabular}
\end{table}

As list in Table \ref{tab:models_overview}, the models selected for evaluation are Vision Transformer (ViT) \cite{vit}, Convolutional Neural Networks Next (ConvNeXt) \cite{convnext}, Convolutions Vision Transformer (CVT) \cite{cvt}, Swin Transformer (Swin) \cite{swin}, Semantic Segmentation with Transformers (SegFormer) \cite{szegedy2013intriguing}, and Residual Networks (ResNet) \cite{resnet}. These models are top-performance sourced from the Huggingface open repository. The publisher has pre-trained these models.
We select the ResNet \cite{resnet} as a baseline. 
Additionally, our experiments employ various CAM-based methods \cite{Grad-cam,Grad-cam++,HiResCAM,xgradcam,layercam} to generate saliency maps for vision models.

\textbf{Computational Cost Analysis:}
Evaluating computational costs is essential to understand the practical implications of deploying XAI techniques in applications.
The assessment covers processing time and energy consumption for different AI models using XAI techniques.

\begin{table}[ht]
\centering
\caption{Analysis of Computational Costs for Vision Models Based on per One Thousand Images}
\label{tab:runtime_comparison_models}
\begin{tabular}{p{1.5cm}p{1.5cm}p{1.5cm}p{1.5cm}p{1.5cm}}
\toprule
Model & Inference Time (s) & Mean XAI Time (s) &  Pipeline Energy (Wh)\\
\midrule
ViT          & 13.35 & 62.50 &  5.58\\
ConvNext     & 5.73  & 41.19 &  2.67 \\
CVT          & 16.91 & 69.76  & 5.86\\
ResNet       & 10.37 & 45.34  & 3.32\\
Swin         & \textbf{32.38} & \textbf{137.24} &  \textbf{10.07}\\
SegFormer    & 13.62 & 65.08 & 5.73\\
\bottomrule
\end{tabular}
\end{table}

Table \ref{tab:runtime_comparison_models} illustrates each model's combined inference time, XAI time, and total energy in the pipelines.
The results indicate significant variation in total processing time and energy consumption across models.
The Swin Transformer shows the highest total process time and energy consumption, indicating a significant computational demand. 
The CVT model consumes the second most energy and runtime.
Conversely, ConvNext has the lowest energy consumption and a relatively shorter total processing time.
This is reasonable because the employed ConvNext model \cite{convnext} is a relatively tiny-sized model. 
This conclusion suggests that CNN-based models, such as ConvNext and ResNet, are still potentially more viable options for applications where energy efficiency and faster processing are crucial.

In summary, these findings indicate significant variations in both time and energy consumption across different models.
XAI consumes significantly higher computational costs compared to model inference.

\textbf{Model Performance Analysis:}
The objective is to evaluate and compare the performance of leading open-source vision models from the Huggingface repository, applying a range of adversarial perturbations.

\begin{figure}[h]
\centering
\includegraphics[width=0.82\linewidth]{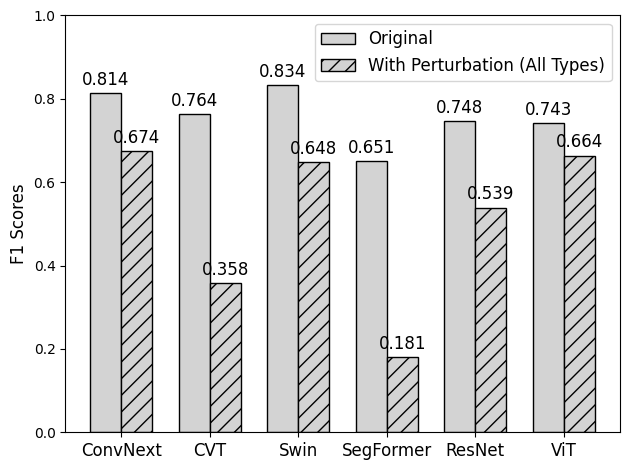}
\caption{Vision Models F1 Scores: Original Dataset and Adversarial Dataset Averages.}
\label{fig:f1}
\end{figure}

Figure \ref{fig:f1} displays the F1 scores for six computer vision models under both original and adversarial perturbation conditions, emphasizing their performance. 
Initially, all models exhibit high performance on the original dataset.
However, adversarial perturbations have varying effects on the performance of each model.
Models indicate performance degradation under adversarial perturbations. 
Specifically, CVT and SegFormer show obvious vulnerability against adversarial perturbations.

\textbf{Model Robustness Analysis:}
This analysis evaluates model robustness against different levels of adversarial perturbations.
The aim is to quantitatively evaluate and compare the robustness by assessing their ability to maintain prediction accuracy.
To analyze changes in prediction distributions under adversarial perturbations and further assess model robustness, we employ the introduced Kolmogorov-Smirnov (K-S) statistic.
Initially, we extract the probability values from the model inference of the original dataset.
Next, we extract prediction probabilities for each perturbed dataset and compute the K-S statistic.
A higher K-S value indicates a greater shift in the model's outputs, implying a significant reduction in robustness.

\begin{figure}[ht]
\centering
\includegraphics[width=0.85\linewidth]{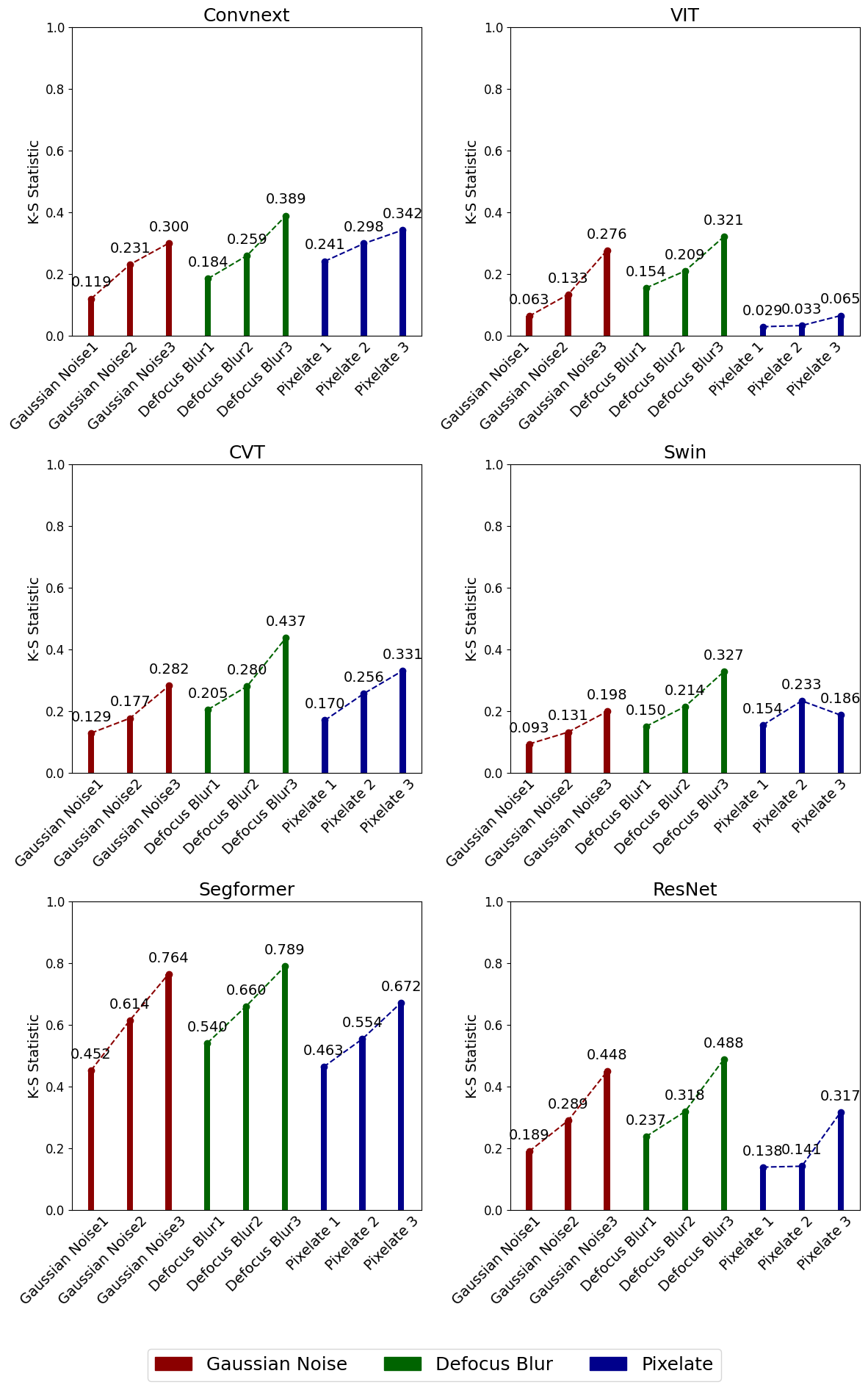}
\caption{Comparison of K-S Statistics to Assess Model Robustness under Three Levels and Types of Perturbations (0: Identical, 1: Highly Divergent).}
\label{fig:resilience}
\end{figure}
 
Figure \ref{fig:resilience} shows the K-S values offer perspectives on model robustness. 
SegFormer consistently exhibits higher K-S values across all perturbations, indicating significant prediction shifts under adversarial attacks.
This suggests vulnerability in maintaining prediction consistency against such perturbations.
Conversely, models such as ViT and Swin exhibit lower K-S values, implying more robust performance under adversarial conditions.
The bar plots in Figure \ref{fig:resilience} also display K-S statistic values for each perturbation type (Gaussian Noise, Defocus Blur, Pixelate) across three levels (1, 2, 3).
Each bar's height indicates the deviation of the model's prediction distribution from its original, caused by a specific perturbation.
Different colors for each perturbation type enhance visual distinction.

As a result, this quantitatively evaluates the robustness of various models under diverse adversarial perturbations. The Vision Transformer and Swin Transformer models exhibit relatively higher robustness.

\textbf{Explanation Deviation Analysis:}
The objective is to evaluate the impact of adversarial perturbations on the explanation deviation.
The utilization of saliency maps to annotate images assists developers in identifying the reason for model inaccuracies, thereby enhancing model performance. 

\begin{figure}[ht]
\centering
% Include your figure file here
\includegraphics[width=1\linewidth]{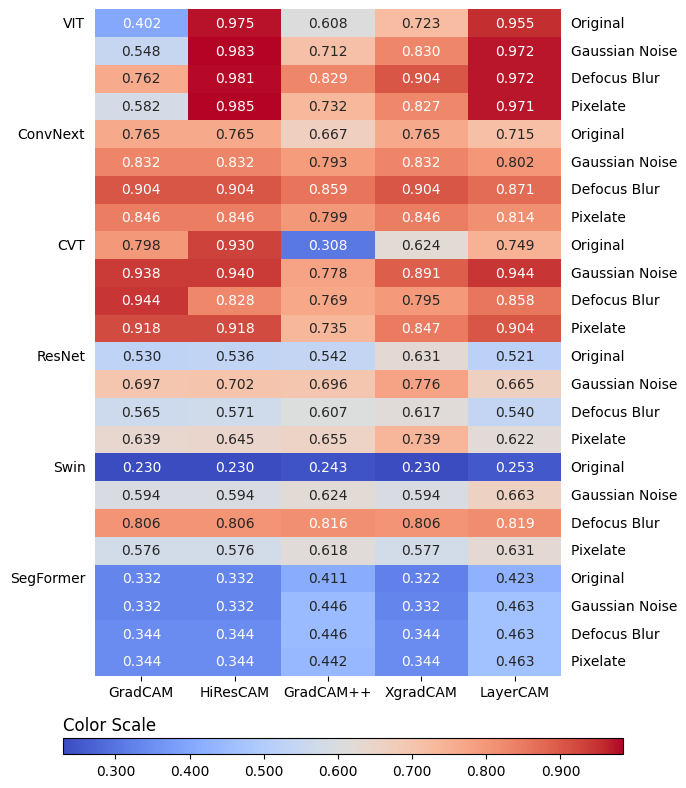}
\caption{Heatmaps Illustrating Median Prediction Change Percentage for Original and Adversarial Perturbed Images. Lower Values Indicate Better Explanation Deviation.}
\label{fig:explainability_heatmaps}
\end{figure}

Figure \ref{fig:explainability_heatmaps} shows prediction change percentages on a heatmap.
The explanation deviation attributes can be calculated as one minus the value shown in the figure. 
The figure is 3-dimensional: the left rows label six vision models, while the right rows specify the original dataset and three types of adversarial perturbation. The labels at the bottom represent various XAI methods. 
The value in each cell reflects the summarized percentile of prediction value changes observed between the original input and the explanation inputs. 
Lower values mean the model's inference is less affected by irrelevant feature masking, indicating XAI makes a closer approximation of relevant features.
Consequently, a lower summarized prediction change percentage values in Figure \ref{fig:explainability_heatmaps}, correlates with better XAI explanation deviation. 

Significant variations are observed across different model-XAI combinations.
The results reveal that the Swin Transformer model maintains consistent explanation deviation across varied XAI methods and adversarial perturbation types, as confirmed through extensive cross-validation.

\textbf{Explanation Resilience Analysis:}
The impact of adversarial perturbations in XAI scenarios remains an under-explored question.
This analysis seeks to determine if adversarial perturbations contribute to misleading explanations.

Figure \ref{fig:explainability_heatmaps} shows the prediction change percentages under adversarial perturbations, which are used to calculate explanation resilience attributes.
Additionally, Table \ref{tab:explanation_explainability_models} offers a concise summary of the explanation resilience results. 
The smaller the resilience value, the better the explanation can resist the impact of adversarial attacks.

\begin{table}[ht]
\centering
\caption{Summary of Orginal Deviation, Adversarial Deviation and Explanation Resilience for Vision Models}
\label{tab:explanation_explainability_models}
\begin{tabularx}{\linewidth}{ 
  >{\hsize=0.3\hsize}X 
  >{\hsize=0.3\hsize}X 
  >{\hsize=0.3\hsize}X 
  >{\hsize=0.2\hsize}X 
}
\toprule
Model & Original & Adversarial & Resilience \\
\midrule
VIT        & 0.267 & 0.160 & 0.107\\
ConvNext   & 0.265 & 0.155 & 0.110\\
CVT        & 0.310 & 0.133 & 0.177\\
ResNet     & 0.448 & 0.351 & 0.097\\
Swin       & 0.763 & 0.327 & \textbf{0.436}\\
SegFormer  & 0.636 & 0.614 & 0.022 \\
\bottomrule
\end{tabularx}
\smallskip 
\end{table}

% The results uncover varying degrees of explanation resilience across different models and XAI methods. The findings indicate that some models maintain a consistent level of explanation deviation, while others demonstrate a notable decrease when faced with adversarial perturbations.

% \begin{mdframed}[innertopmargin=5pt,linecolor=black,roundcorner=5pt,backgroundcolor=yellow!20] 
% \end{mdframed}

The analysis shows that the explanation resilience attribute does not align with performance and robustness attributes.
The findings reveal that models, such as Swin Transforme, retain consistent explanation deviation, whereas show a significant decrease when encountering adversarial perturbations.

\subsection{The Assessment Scenario of Tabular Models}
In the case of structured data, we present the experimental results for all quality attributes.

\textbf{Process Configuration:}
Our study evaluates tabular models using datasets from various domains.
COMPAS \cite{COMPAS} Recidivism Risk Score Data and Analysis dataset is instrumental in assessing the predictive accuracy of recidivism, offering a rich source of sensitive real-world data. 
RT-IoT2022 \cite{rt-iot}, the 2022 Real-Time Internet of Things (IoT) dataset \cite{rt-iot} for cybersecurity threats case, is a collection of network traffic data. 
PriceRunner \cite{productcluster} Product Classification and Clustering dataset provides a scenario in the e-commerce domain.

The TabTransformer \cite{tabtransformer}, TabNet \cite{tabnet}, and FT Transformers \cite{fttransformer} are employed as the models for tabular data.
The TabTransformer \cite{tabtransformer}, a model inspired by the Transformer architecture, is adapted for tabular data by encoding categorical features into embeddings. 
TabNet \cite{tabnet} utilizes sequential attention to choose features for each decision step. 
The FT Transformer \cite{fttransformer} optimizes the transformer model to handle numerical features. 
In terms of XAI methods, Mean Centroid Prediff \cite{Trustworthy} and SHAP \cite{lundberg2017unified} are applied.

\textbf{Computational Cost Analysis:}
For the computational cost analysis, we test the time and energy consumption for the three model inference, Mean centroid prediff \cite{Trustworthy}, and SHAP \cite{lundberg2017unified} method. Table \ref{tab:rfc_efficiency} presents the recorded time and energy consumption for every thousand samples using the algorithms.

\begin{table}[h]
\centering
\caption{Analysis of Computational Costs for Tabular Models Based on per Thousand Rows}
\label{tab:rfc_efficiency}
\begin{tabular}{lcc}
\toprule
Model and XAI & Time (s) & Energy (Wh) \\
\midrule
TabTransformer & 16.04 & 0.49 \\
TabNet & 19.39 & 0.56\\
FT Transformers & 15.62 & 0.47\\
Mean Centroid Prediff & 1112.28 & 53.87 \\
SHAP & 768.13 & 26.74 \\
\bottomrule
\end{tabular}
\end{table}

The costs for the three transformer-based tabular models are comparable.
However, XAI takes significantly higher computational costs than model inference. It is observed that the Mean Centroid Prediff \cite{Trustworthy} has higher time and energy consumption compared to SHAP \cite{lundberg2017unified}.

\textbf{Model Performance and Robustness Analysis:}
We evaluated the performance of various transformer-based tabular models \cite{tabtransformer,tabnet,fttransformer} against adversarial perturbations. 
The performance metrics before and after the adversarial perturbations are shown in Table \ref{tab:performance_explainability}.

% \begin{table}[ht]
% \centering
% \caption{COMPAS Case Model Performance Comparison.}
% \begin{tabularx}{\linewidth}{X|>{\centering\arraybackslash}X>{\centering\arraybackslash}X>{\centering\arraybackslash}X|>{\centering\arraybackslash}X>{\centering\arraybackslash}X>{\centering\arraybackslash}X}
% \hline
% & \multicolumn{3}{c|}{\textbf{Original Data}} & \multicolumn{3}{c}{\textbf{Adversarial Data}} \\
% \cline{2-7} 
% \textbf{Model} & \textbf{Precision} & \textbf{Recall} & \textbf{F1 Score} & \textbf{Precision} & \textbf{Recall} & \textbf{F1 Score} \\
% \hline
% Random Forest & 0.875 & 0.929 & 0.901 & 0.898 & 0.801 & 0.847 \\
% LightGBM & 0.874 & 0.944 & 0.908 & 0.869 & 0.907 & 0.888 \\
% XGBoost & 0.874 & 0.928 & 0.900 & 0.832 & 0.964 & 0.893 \\
% \hline
% \end{tabularx}
% \label{table:model_performance_comparison}
% \end{table}

% \begin{table}[ht]
% \centering
% \caption{COMPAS Case Model Performance Comparison on Original and Adversarial Data}
% \label{table:model_performance_comparison}
% \begin{tabular}{lccc}
% \toprule
% \textbf{Model} & \textbf{F1 Score (O)} & \textbf{F1 Score (A)} & \textbf{K-S Statistic} \\
% \midrule
% Random Forest & 0.901 & 0.847 & 0.067 \\
% LightGBM      & 0.908 & 0.888 & 0.054 \\
% XGBoost       & 0.900 & 0.893 & 0.024 \\
% \bottomrule
% \end{tabular}
% \end{table}

\begin{table}[ht]
\centering
\caption{Analysis of Model Performance and XAI Deviation in Tabular Scenarios}
\label{tab:performance_explainability}
\resizebox{\linewidth}{!}{%
\begin{tabular}{lccccc}
\toprule
\multirow{2}{*}{\textbf{Dataset}} & \multirow{2}{*}{\textbf{Model}} & \multicolumn{2}{c}{\textbf{Model Performance}} & \multicolumn{2}{c}{\textbf{XAI Deviation}} \\
\cmidrule(lr){3-4} \cmidrule(lr){5-6}
                                  &                                 & \textbf{Original} & \textbf{Adv. Changes} & \textbf{Original} & \textbf{Adv. Changes} \\
\midrule
\multirow{3}{*}{COMPAS}           & TabTransformer                  & 0.683             & -0.118               & 0.965              & -0.105 \\
                                  & TabNet                          & 0.674             & -0.114               & 0.989              & -0.087 \\
                                  & FT Transformers                 & 0.690             & -0.103               & 0.965              & -0.071 \\
\midrule
\multirow{3}{*}{RT-IoT}           & TabTransformer                  & 0.921             & -0.435               & 0.987              & -0.060 \\
                                  & TabNet                          & 0.883             & -0.364               & 0.985              & -0.078 \\
                                  & FT Transformers                 & 0.950             & -0.603               & 0.984              & -0.088 \\
\midrule
\multirow{3}{*}{PriceRunner}      & TabTransformer                  & 0.994             & -0.175               & 0.977              & -0.063 \\
                                  & TabNet                          & 0.991             & -0.179               & 0.973              & -0.062 \\
                                  & FT Transformers                 & 0.997             & -0.174               & 0.973              & -0.049 \\
\bottomrule
\end{tabular}%
}
\end{table}

The adversarial perturbation causes the model performance to decrease. Among the tests, the performance results decrease remarkably, especially for the RT-IoT dataset, which primarily consists of numerical data derived from sensors. 

% \subsubsection{Scenario Two: Model Performance Analysis}

% The model's performance was evaluated, with adversarial perturbation.
% The performance metrics show minimal change: Precision decreased slightly from 0.580 to 0.574, Recall from 0.605 to 0.602, and F1 Score from 0.592 to 0.588 under adversarial conditions. 
% This is due to the applied adversarial attack designed for fooling the XAI \cite{fooling}. The only sensitive feature ``race" is perpetuated.

% \begin{table}[ht]
% \centering
% \caption{Model Performance Metrics Under Standard and Adversarial Conditions}
% \label{tab:model_performance_adversarial}
% \begin{tabular}{lccc}
% \toprule
% Condition & Precision & Recall & F1 Score \\
% \midrule
% Standard & 0.580 & 0.605 & 0.592 \\
% Adversarial & 0.574 & 0.602 & 0.588 \\
% \bottomrule
% \end{tabular}
% \end{table}

\textbf{Explanation Deviation and Resilience Analysis:}
We employ the Mean Centroid Prediff \cite{Trustworthy} and SHAP \cite{lundberg2017unified} methodologies to calculate feature importance values. These explanation deviations are also shown in Table \ref{tab:performance_explainability}.
The average adversarial decrement in explanation deviation for Mean Centroid Prediff stands at 0.047, in contrast to a 0.099 reduction for SHAP.
This reveals that Mean Centroid Prediff exhibits superior resilience compared to SHAP; however, it comes at the cost of increased computational demands.

\subsection{Conclusions and Insights on Research Questions}
\begin{figure}[ht]
\centering
\includegraphics[width=\linewidth]{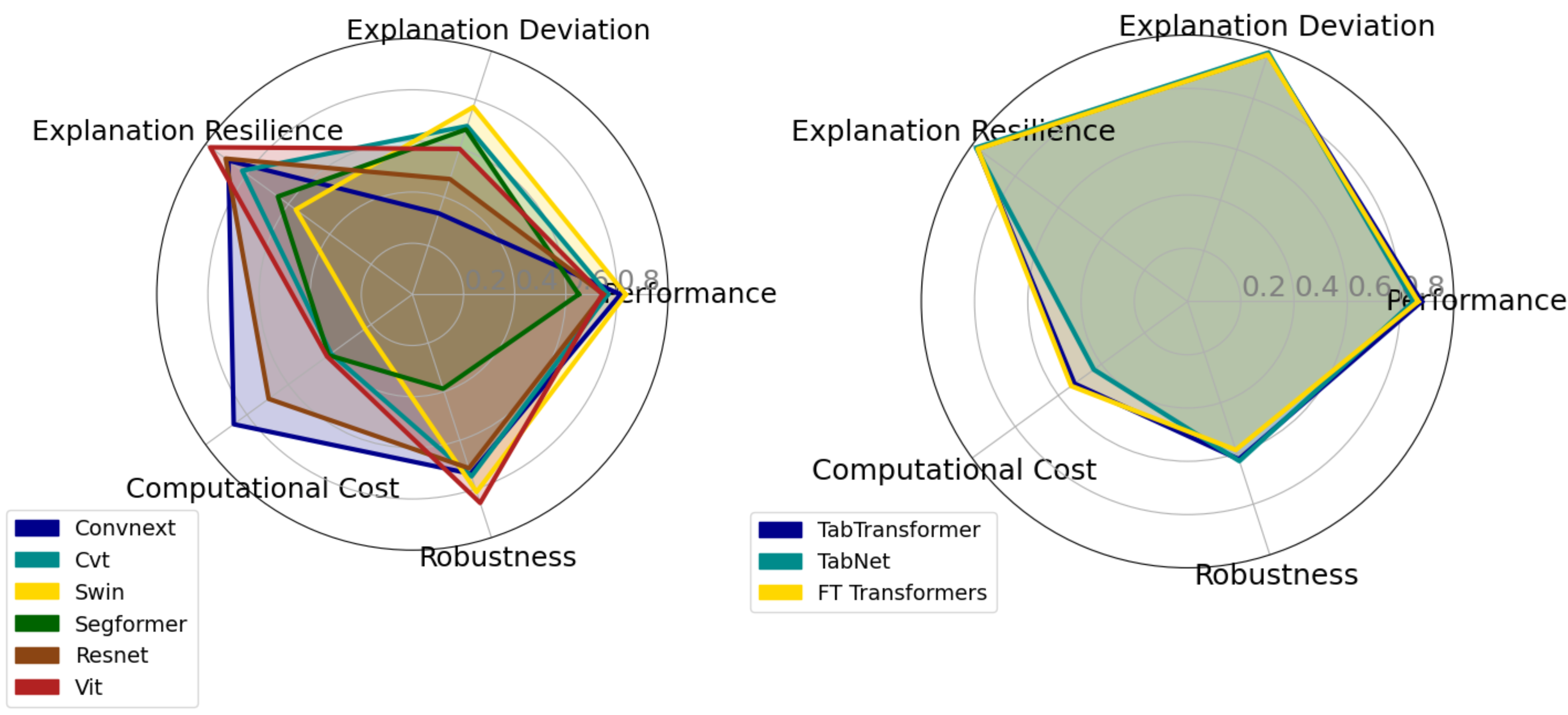}
\caption{Comprehensive Overview of Multiple Quality Attributes Assessment for the Selected Models.}
\label{fig:radar}
\end{figure}

The five quality attributes are designed to comprehensively evaluate the open-source models.
Figure \ref{fig:radar} presents a radar chart visualizing the comparative analysis of selected vision and tabular models across the quality attributes.
It is noted that the chart's values have been normalized, where each value closer to one indicates better the attributes.
This evaluation aims to support engineers in selecting and developing explainable models in AI-enabled software.

\textbf{Response to RQ1:}
The evaluation results show variability in the explainability metrics of XAI methods across various models.
The scenario studies encompassed various models, ranging from CNN-based to Transformer-based and from vision to tabular data.
When combined with various models, XAI methods offer diverse explanation utilities, as detailed in Figure \ref{fig:explainability_heatmaps}.
This variability highlights the importance of assessing compatibility between models and XAI methods.
This study introduces a cloud-based XAI service that automates the evaluation pipeline, thereby streamlining the assessment of explainability metrics across diverse models and methods.
By providing ease of customization, this service enables microservice to be reusable and the pipeline reproducible.

\textbf{Response to RQ2:}
The investigation into the computational costs of XAI shows a notable burden, especially when compared to the costs for model inferences. 
The experimental XAI pipelines, as summarized in Table \ref{tab:models_deviation_energy}, show a correlation between a model's explanation deviation and its computational demands.
\begin{table}[h]
\centering
\caption{Explanation Deviation and Energy Consumption of the Selected Models}
\label{tab:models_deviation_energy}
\begin{tabular}{lcc}
\toprule
\textbf{Models(Vision/Tabular)} & \textbf{Explanation Deviation $\downarrow$} & \textbf{Energy (Wh) $\downarrow$} \\
\midrule
Swin & 0.770 & 10.07 \\
CVT & 0.692 & 5.86 \\
SegFormer & 0.678 & 5.73 \\
ViT & 0.598 & 5.58 \\
ResNet & 0.474 & 3.32 \\
ConvNeXt & 0.333 & 2.67 \\
\midrule
TabNet & 0.983 & 0.56 \\
TabTransformer & 0.977 & 0.49 \\
FT Transformers & 0.974 & 0.47 \\
\bottomrule
\end{tabular}
\end{table}
The Swin Transformer model, offering the highest model performance and explanation deviation, also incurs the highest computational costs, as indicated by processing times and energy consumption metrics.
This correlation showcases a trade-off between achieving desirable levels of explanation deviation and the associated computational costs, suggesting the need to evaluate the model-XAI combination and optimize this balance.

\textbf{Response to RQ3:}
The experiments demonstrate that adversarial perturbations significantly affect both model performance and explanations, as detailed in Table \ref{tab:model_robustness}. 
% This analysis showcases the necessity for a comprehensive assessment of the model, emphasizing the need to balance the set of quality attributes in application scenarios.

\begin{table}[ht]
\centering
\caption{Model Performance and Explanation Deviation Changes with Adversarial Attacks, Results from 108 XAI Pipelines}
\label{tab:model_robustness}
\begin{tabular}{@{}lcc@{}}
\toprule
\multicolumn{1}{c}{\textbf{Attribute}} & \textbf{Significant ($p > 0.05$)} & \textbf{Non significant ($p \leq 0.05$)} \\ 
\midrule
Performance & 88.89\% & 11.11\% \\
Deviation & 69.44\% & 30.56\% \\
\bottomrule
\end{tabular}
\end{table}

In addition to p-value, we employ Cliff's Delta analysis \cite{cliff2014ordinal} to compare the impact of adversarial attacks. 
A Cliff's Delta of 0.129 for model performance suggests models are more accurate without adversarial attacks.
A Cliff's Delta of 0.428 for explanation deviation indicates greater accuracy of XAI results without adversarial attacks.
Our cloud-based XAI service offers an automated framework for assessing model-XAI combinations and acquiring quality attributes via executing designed pipelines.

\section{Conclusion}
\label{chap:conclusion}

This study proposes an XAI service framework designed to streamline operational complexities in XAI evaluation. 
The framework provides API and SDK interfaces for operation and enables deployment on cloud platforms.
It facilitates data preprocessing, encompassing the processing of datasets, data transformations, and the application of adversarial perturbations. 
The service encapsulates AI models and XAI methods, enabling flexible combinations arranged by the task configuration.
We develop and implement evaluation pipelines to assess five key quality attributes: computational cost, performance, robustness, XAI deviation, and XAI resilience.

The findings lead us to conclude with three research questions. First, we observed the variability in metrics results across models and XAI methods. This shows the necessity of evaluation before selecting the XAI method for AI models. This ensures effective explanations are provided to stakeholders. Second, our analysis summarizes the results of computational cost and the explanation metric. High explanation deviation often requires the expense of increased computational resources. Finally, we demonstrate that adversarial perturbations affect both the model and the XAI, thereby emphasizing the importance of incorporating robust model and XAI methods.
These multidimensional quality attributes guide researchers and practitioners in making informed decisions in AI-based software development and deployment.

\bibliographystyle{IEEEtran}
\bibliography{reference.bib}

\end{document}